\documentclass{article}%
\usepackage{amsmath}
\usepackage{amsfonts}
\usepackage{amssymb}
\usepackage{graphicx}%
\providecommand{\U}[1]{\protect\rule{.1in}{.1in}}
\newcommand{\Dslash}[1] {
\setbox0=\hbox{$#1$}     \dimen0=\wd0        \setbox1=\hbox{/} \dimen1=\wd1
\ifdim\dimen0>\dimen1         \rlap{\hbox to \dimen0{\hfil/\hfil}}       #1
\else           \rlap{\hbox to \dimen1{\hfil$#1$\hfil}}       /         \fi
}

\newcommand{\ba}{\begin{array}}
\newcommand{\ea}{\end{array}}

\newcommand{\be}[1]{
\begin{eqnarray}\label{#1}}
\newcommand{\ee}{\end{eqnarray}}
\newcommand{\bea}{\begin{eqnarray}}
\newcommand{\eea}{\end{eqnarray}}

\newcommand{\ns}{\Dslash{n}}
\newcommand {\nbs}{\Dslash{\bar n}}
\newcommand{\nbn}{\frac{\nbs\ns}{4}}

\newcommand{\ALL}{A_{\text{LL}} }
\newcommand{\KLL}{K_{\text{LL}} }
\newcommand{\KLS}{K_{\text{LS}} }
\def\Dsl[#1]#2{#1\hskip-#2pt/} 

\begin{document}

\begin{center}
{\Large  Wide angle Compton scattering on the proton: study of power suppressed corrections}

\vspace*{1cm}

N. Kivel\footnote{
On leave of absence from St.~Petersburg Nuclear Physics Institute,
188350, Gatchina, Russia} and  
M. Vanderhaeghen
\\[3mm]
 { \it Helmholtz Institut Mainz, Johannes Gutenberg-Universit\"at, D-55099 Mainz, Germany
 \\
 Institut f\"ur Kernphysik, Johannes Gutenberg-Universit\"at, D-55099 Mainz, Germany
 }

\vspace*{1cm}
\end{center}

\begin{abstract}

We  study the wide angle Compton scattering process on a proton within the  soft collinear factorization (SCET) framework. The main
purpose of this work is to estimate the effect due to certain power suppressed corrections.  We consider 
all possible kinematical power corrections and also  include  the subleading amplitudes describing the scattering with
nucleon  helicity flip. Under certain assumptions we present a  leading-order  factorization formula for these amplitudes which includes
the hard- and soft-spectator contributions.  We apply the formalism and perform  a phenomenological analysis of the cross section and asymmetries in the wide angle Compton scattering on a proton.
We assume that in the relevant kinematical region where $-t,-u>2.5$~GeV$^{2}$  the dominant  contribution is provided by the 
soft-spectator  mechanism. The hard coefficient functions  of the corresponding SCET operators are taken in the 
leading-order approximation.    The analysis of  existing cross section data 
shows that the contribution of the  
 helicity flip amplitudes to this observable   is quite 
small and   comparable with other expected  theoretical uncertainties.
We also show predictions for double polarization observables for which experimental 
information exists.

\end{abstract}

\vspace*{1cm}

\section{Introduction}
Wide angle Compton scattering (WACS) on a proton  is  one of the most basic  processes within  the broad class of  hard exclusive reactions aimed at studying the partonic structure of the nucleon.   
  The first data for the differential cross section of this process  has already been obtained long time ago \cite{Shupe:1979vg}.  New  and more precise measurements were carried out at JLab \cite{Danagoulian:2007gs}.  
  Double polarization observables for a polarized photon beam and by measuring the polarization of the recoiling proton
  were also  measured at Jefferson Lab (JLab) \cite{Hamilton:2004fq}.  New measurements of  various observables  at higher energies 
  are planned at the new JLAB 12~GeV facility,  see e.g. \cite{PR1214003}.    
  
  The  asymptotic limit of the WACS cross section, as   predicted by QCD factorization, has been  studied  in many  theoretical works
  \cite{Kronfeld:1991kp, Vanderhaeghen:1997my, Brooks:2000nb, Thomson:2006ny}.  It was found that  the leading-twist  contribution described by the hard two-gluon exchange between three collinear quarks  predicts much smaller   cross sections than is observed in  experiments.  One of the most promising  explanations  of this problem  is that  the kinematical region  of the existing data is still far away from  the asymptotic limit where the hard two-gluon exchange mechanism is predicted  to dominate.  Hence one needs to develop an  alternative  theoretical approach which is more  suitable for  the kinematic range of existing experiments. 
  
Several  phenomenological considerations, including the  large value of the asymmetry $\KLL$ \cite{Hamilton:2004fq} indicate that the dominant  contribution  in the relevant kinematic range can be provided by the so-called soft-overlap mechanism.  In this case the underlying quark-photon scattering  is described by the handbag diagram with one active quark while the  other spectator quarks are assumed to be soft.  Various models have been considered  in order to implement such scattering  picture within a theoretical framework:   diquarks  Refs.\cite{Kroll:1990kh}, GPD-models \cite{Radyushkin:1998rt, handbag1,handbag2,handbag3}  and  constituent quarks \cite{Miller:2004rc} .  

An attempt to develop a  systematic approach  within the soft collinear effective theory (SCET) framework was discussed in Refs.\cite{Kivel:2012vs,Kivel:2013sya}.  The  description  can be considered  as  a natural extension of the collinear factorization to the case with  soft spectators.  In our previous works  the factorization  of the three leading power amplitudes has been studied  and a  phenomenological analysis was made.  The three  amplitudes  describing  Compton  scattering which involve a nucleon helicity-flip are power suppressed and they were neglected in our previous analysis.   In the present work we   want  to  include   these amplitudes into our description, together with all  kinematical power corrections.  For that purpose we discuss  the  factorization   of  helicity-flip amplitudes  assuming that  it can be described as  a sum of  hard- and soft-spectator contributions.  We show that the corresponding soft contributions  are described by the appropriate  subleading so-called  SCET-I operators.  
As a first step towards a proof 
 of the factorization  we   restrict our attention  only to the relevant  operators which  appear in the leading-order  approximation in $\alpha_{s}$.    Assuming  that  such soft contributions are dominant  we  estimate their possible numerical impact  on the cross section and asymmetries.  

Our  work is organized as follows.  In Sec.\ref{kin} we shortly describe the kinematics, amplitudes, cross sections and  asymmetries. In Sec.\ref{fact} we  discuss the factorization scheme for  the subleading amplitudes, describe the suitable SCET-I operators and their matrix elements.  We  also compute  the corresponding  leading-order coefficient functions and  provide the resulting expressions for the amplitudes.  Sec.\ref{phenom} is devoted to a phenomenological analysis and in Sec.\ref{dis} we summarize our conclusions.

\section{Kinematics and observables}
\label{kin}
In this paper we follow the  notations introduced in Ref.\cite{Kivel:2013sya}. 
For convenience,    we briefly summarize   the most important details.  
In our  theoretical consideration we will use the Breit  frame where {\it in} and {\it out}  nucleons (with momenta $p$ and $p'$ respectively)  move along
the $z$-axis and $ p_{z}=- p_{z} '$.  Using  the auxiliary light-like vectors 
\begin{equation}
\ n=(1,0,0,-1),~\bar{n}=(1,0,0,1),~\ (n\cdot\bar{n})=2,
\end{equation}
 the light-cone expansions of the momenta can be written as follows
\begin{equation}
p=W\frac{\bar{n}}{2}+\frac{m^{2}}{W}\frac{n}{2},~p^{\prime}=\frac{m^{2}}{W}\frac{\bar{n}}{2}+W\frac{n}{2},
 \label{mom1}
\end{equation}
where  $m$ is nucleon mass and the convenient variable $W$ can be expressed  through the momentum transfer $t$ as 
\begin{equation}
W=m\left (\sqrt{\frac{-t}{4m^{2}}}+\sqrt{1+\frac{-t}{4m^{2}}}\right).
\end{equation} 
The photon momentum reads
\bea
q=(q\cdot  n)\frac{\bar{n}}{2}+(q\cdot \bar n)\frac{n}{2}+q_{\bot},
\eea
with
\begin{equation}
(q\cdot  n)=-\frac{\left(  u-m^{2}\right)  +\kappa(s-m^{2})}{W(1-\kappa^{2})},~
(q\cdot \bar n)=\frac{s-m^{2}+\kappa\left(  u-m^{2}\right)}{W(1-\kappa^{2})},
\end{equation}
where $\kappa=m^{2}/W^{2}$.  In the limit  $s\sim -t\sim -u \gg m^{2}$  these expressions can be simplified neglecting  the power suppressed contributions 
\bea
 p\simeq W\frac{\bar{n}}{2},~p^{\prime}\simeq W\frac{n}{2}, \
q\simeq\frac{-u}{W} \frac{\bar{n}}{2}+\frac{s}{W}\frac{n}{2}+q_{\bot},~\ q^{\prime}\simeq \frac{s}{W}\frac{\bar{n}}{2}+\frac{-u}{W} \frac
{n}{2}+q_{\bot}.
\eea
where we assume that $W\simeq \sqrt{-t}$. 

For the amplitude we borrow the  parametrization from Ref.\cite{Babusci:1998ww}
\bea
M^{\gamma p\rightarrow \gamma p}&=&-e^{2}\,  \varepsilon^{\ast\mu}( q^{\prime})\varepsilon^{\nu}(q) 
\bar{N}(p^{\prime})\mathcal{A}^{\mu\nu} N(p),
\\
\mathcal{A}^{\mu\nu}&=&\left\{-\mathcal{T}_{12}^{\mu\nu}\left(  T_{1}+\Dsl[K]{7}~T_{2}\right)  -\mathcal{T}_{34}^{\mu\nu
}\left( T_{3}+\Dsl[K]{7}~T_{4}\right)
 +\mathcal{T}_{5}^{\mu\nu}i\gamma_{5}~T_{5}%
+\mathcal{T}_{6}^{\mu\nu}~i\gamma_{5}\Dsl[K]{7}~T_{6}\right\},
\label{defM}
\eea
where  $e$ denotes the electromagnetic charge of the proton, $N(p)$ is the nucleon spinor.  In Eq.(\ref{defM}) we introduced  the orthogonal  tensor  structures
\begin{align}
  \mathcal{T}_{12}^{\mu\nu}=-\frac{P^{\prime\mu}P^{\prime\nu}}{P^{\prime2}},\quad   
 \mathcal{T}_{34}^{\mu\nu}=\frac{N^{\mu}N^{\nu}}{N^{2}},\quad
  \mathcal{T}_{5}^{\mu\nu}=\frac{P^{\prime\mu}N^{\nu}-P^{\prime\nu}N^{\mu}}{P^{\prime2}K^{2}},\quad
\mathcal{T}_{6}^{\mu\nu}=\frac{P^{\prime\mu}N^{\nu}+P^{\prime\nu}N^{\mu}}{P^{\prime2}K^{2}},
 \label{T1-6}
\end{align}%
with%
\bea
P=\frac{1}{2}(p+p^{\prime}),~K=\frac{1}{2}(q+q^{\prime}),~P^{\prime}=P-K\frac{(P\cdot K)}{K^{2}},
N_{\mu}=\varepsilon_{\mu\alpha\beta\gamma}P^{\alpha}\frac{1}
{2}(p-p^{\prime})^{\beta}K^{\gamma}.
\eea
The scalar amplitudes $T_{i}\equiv T_{i}(s,t)$ are  functions of the Mandelstam variables.

The analytical expressions for various observables can also be found in Ref.\cite{Babusci:1998ww}. 
In our consideration it will be convenient  to redefine two helicity-flip amplitudes  as
\bea
\bar T_{1}=T_{1}+\frac{m(s-u)}{4m^{2}-t}T_{2},\quad \bar T_{3}= T_{3}+\frac{m(s-u)}{4m^{2}-t}T_{4},
\label{defT1b}
\eea
The reason for such redefinition will be clarified later.  The cross section  reads 
\begin{equation}
\frac{d\sigma}{dt}=\frac{\pi\alpha^{2}}{(s-m^{2})^{2}}~W_{00},
\end{equation}
with%
\begin{align}
W_{00}    =
\frac{(m^{4}-u s)(-t)}{(4m^{2}-t)}\left(  \frac12 \left\vert T_{2}\right\vert^{2}+ \frac12  \left\vert T_{4}\right\vert^{2}+\left\vert T_{6}\right\vert ^{2}\right)
+\frac{1}{2}(4m^{2}-t)
\left(  \left\vert \bar T_{1}\right\vert ^{2}+ \left\vert \bar T_{3}\right\vert ^{2}\right) -t\left\vert T_{5}\right\vert ^{2},
\label{W00}
\end{align}
c.f. with Eq.(3.15a) in Ref.\cite{Babusci:1998ww}. 
We also describe the  asymmetries which will be considered  in this work. We are interested in the beam target-asymmetries  with circular photon polarization 
($R, L$). In the case of a  longitudinally polarized nucleon target, the   corresponding asymmetry  $\ALL$   reads  (in c.m.s)
\begin{equation}
\ALL=-\frac{\sigma_{z}^{R}-\sigma_{z}^{L}}{\sigma_{z}^{R}+\sigma_{z}^{L}}=-\frac{C_{z}^{K}W_{12}^{+}+C_{z}^{Q}W_{22}^{+}}{W_{00}}.
\end{equation}
Two further asymmetries describe the correlations of the recoil polarization  with the polarization of the photons:
\bea
\KLL&=&\frac{\sigma_{z'}^{R}-\sigma_{z'}^{L}}{\sigma_{z'}^{R}+\sigma_{z'}^{L}}=\frac{C_{z'}^{K}W_{12}^{-}+C_{z'}^{Q}W_{22}^{-}}{W_{00}},
\\
 \KLS&=& \frac{\sigma_{x'}^{R}-\sigma_{x'}^{L}}{\sigma_{x'}^{R}+\sigma_{x'}^{L}}=\frac{C_{x^{\prime}}^{K}W_{12}^{-}+C_{x^{\prime}}^{Q}W_{22}^{-}}{W_{00}},
\eea
where (for more details see \cite{Babusci:1998ww}) 
\begin{equation}
W_{12}^{\pm}=(4m^{2}-t)(\bar{T}_{3}-\bar{T}_{1})T_{6}^{\ast}\pm t\left(
T_{2}+T_{4}\right) {T}_{5}^{\ast},
\label{W12}
\end{equation}%
\begin{align}
W_{22}^{\pm}  &  =\pm~4m\frac{m^{4}-su}{4m^{2}-t}\left(  T_{2}-T_{4}\right)
{T}^{*}_{6}\pm(s-u)(\bar{T}_{3}-\bar{T}_{1})T_{6}^{\ast}-(s-u)\left(
T_{2}+T_{4}\right) {T}_{5}^{\ast} -4m(\bar{T}_{3}+\bar{T}_{1}){T}_{5}^{\ast}.
\label{W22}
\end{align}
The   coefficients  $C_{i}^{K,Q}$  read
\bea
C_{z}^{K}=-C_{z^{\prime}}^{K}=-\frac{s-m^{2}}{2m}-\frac{t(s+m^{2})}%
{4m(s-m^{2})}, \\
C_{z}^{Q}=C_{z^{\prime}}^{Q}=-\frac{t(s+m^{2})}{4m(s-m^{2})},
\eea
\begin{equation}
C_{x^{\prime}}^{K}=-C_{x^{\prime}}^{Q}=-\frac{\sqrt{-t(m^{4}-su)}}{2(s-m^{2}%
)}.
\end{equation}

\section{Factorization of  the subleading helicity-flip amplitudes $T_{1,3,5}$}
\label{fact}

In Ref.\cite{Kivel:2013sya},  the factorization of the  helicity conserving amplitudes $T_{2,4,6}$ was considered in the SCET framework \cite{Bauer:2000ew,Bauer2000,Bauer:2001ct,Bauer2001,Beneke:2002ph,Beneke:2002ni}. 
The helicity-flip amplitudes are power  suppressed and were neglected.  
In  the current  paper we would like to extend the SCET  analysis and also consider  the subleading  amplitudes  $T_{1,3,5}$.  
Below we are using  the same notation for the SCET fields and  charge invariant combinations  as in Ref.\cite{Kivel:2013sya}.

The factorization of the helicity conserving amplitudes
$T_{2,4,6}$ is described by the sum of the soft- and hard-spectator
contributions.  It is natural to expect that the same general structure  also holds
for the subleading  amplitudes $T_{1,3,5}$. Therefore  we assume 
that the $T$-product of the electromagnetic currents can be presented  as 
\begin{equation}
T\{J^{\mu}(x),J^{\nu}(0)\}=\sum\tilde{C}^{\mu\nu}\ast {\mathcal O}_{I}+\sum O_{n}^{(i)}\ast\tilde
{T}^{\mu\nu}\ast O_{\bar{n}}^{(j)},  \label{TJJ}%
\end{equation}
where ${\mathcal O}_{I}$ denotes  the different  SCET-I operators associated with the soft-spectator contribution and  $O_{n}^{(i)}\ast\tilde
{T}^{\mu\nu}\ast O_{\bar{n}}^{(j)}$ describes the hard-spectator term with the collinear operators $O_{n}^{(i)}\sim \lambda^{i}$, with  $\lambda\sim \sqrt{\Lambda/Q}$  a generic small parameter.  The sums in (\ref{TJJ})  include  all possible operators in  both terms.  
The power counting of the hard-spectator contribution is provided by the collinear operators
\begin{equation}
O_{n}^{(i)}\ast\tilde{T}^{\mu\nu}\ast O_{\bar{n}}^{(j)}\sim\lambda^{i+j}.
\end{equation}
 These operators are
constructed from the collinear quark and gluon fields. The leading-twist operator  is given  by the three quark operator $ O_{n}^{(6)}=\bar \chi^{c}_{n} \bar \chi^{c}_{n}\bar \chi^{c}_{n}$  and is of order $\lambda^{6}$ (twist-3 operator).   In order to describe  the helicity-flip amplitudes one has to include the subleading operators of
order $\lambda^{8}$ (twist-4).  Therefore the helicity-flip amplitudes are
suppressed by at least a factor $\lambda^{14}$
while the leading power  amplitudes are described by the operator $O_{n}^{(6)}\ast
T\ast O_{\bar{n}}^{(6)}\sim\lambda^{12}$.   The explicit calculations of the
hard-spectator part in Eq.(\ref{TJJ}) is  ill defined,  because of
end-point singularities in the collinear convolution integrals, see for instance the calculation of the form factor $F_{2}$ in Ref.\cite{Belitsky:2002kj}. 
Only the sum of the soft- and hard-spectator contributions in Eq.(\ref{TJJ})  provides a well defined result. 
 
 The soft-spectator contribution is described by  the first term on the \textit{rhs} of Eq.(\ref{TJJ}) 
 where the operators ${\cal O}_{I}$ are constructed from the hard-collinear  fields
in SCET-I.  In Ref.\cite{Kivel:2013sya}  it was  shown that for the leading power contribution this operator reads    
\begin{equation}
{\cal O}_{I}=O^{\sigma}=\sum e_{q}^{2}\left\{  \bar{\chi}_{n}^{q}\gamma_{\bot}^{\sigma}%
\chi_{\bar{n}}^{q}-\bar{\chi}_{\bar{n}}^{q}\gamma_{\bot}^{\sigma}\chi_{n}%
^{q}\right\}  \sim\mathcal{O}(\lambda^{2}). \label{Osgm}%
\end{equation}
The matrix element of this operator gives  only  the helicity conserving amplitudes
\begin{equation}
\left\langle p^{\prime}\right\vert O^{\sigma}\left\vert p\right\rangle _{\text{{\tiny SCET}}}=\bar{N}_{n}\gamma_{\bot}^{\sigma}N_{\bar{n}%
}~\mathcal{F}_{1}(t),
\label{defF1}
\end{equation}
where
\begin{equation}
\bar{N}_{n}=\bar{N}(p^{\prime})\nbn,  {N}_{\bar n}=\nbn  N(p).
\label{Nn}
\end{equation}
  Hence  in order to describe the soft-spectator
contribution of the helicity-flip amplitudes   we need the subleading
operators.  A similar situation also holds for the  proton form factors $F_{1}$ and $F_{2}$ see {\it e.g.}  Ref.\cite{Kivel:2010ns}.

The matrix element of the required subleading operator must  describe  the
chiral-odd Dirac structures appearing in the amplitudes
\begin{equation}
\left\langle p^{\prime}\right\vert \mathcal{O}_{I}\left\vert p\right\rangle
_{\text{{\footnotesize SCET}}}=\bar{N}_{n}1N_{\bar{n}}~A~\mathcal{+~}\bar
{N}_{n}i\gamma_{5}N_{\bar{n}}~B,
\label{meOsub}
\end{equation}
where $A$ and $B$ are some scalar SCET-I amplitudes.  From Eq.(\ref{meOsub})  it follows that the SCET operator $\mathcal{O}_{I}$ can only have an even
number of the transverse Lorentz indices.  

\begin{figure}[ptb]%
\centering
\includegraphics [width=4.5in]{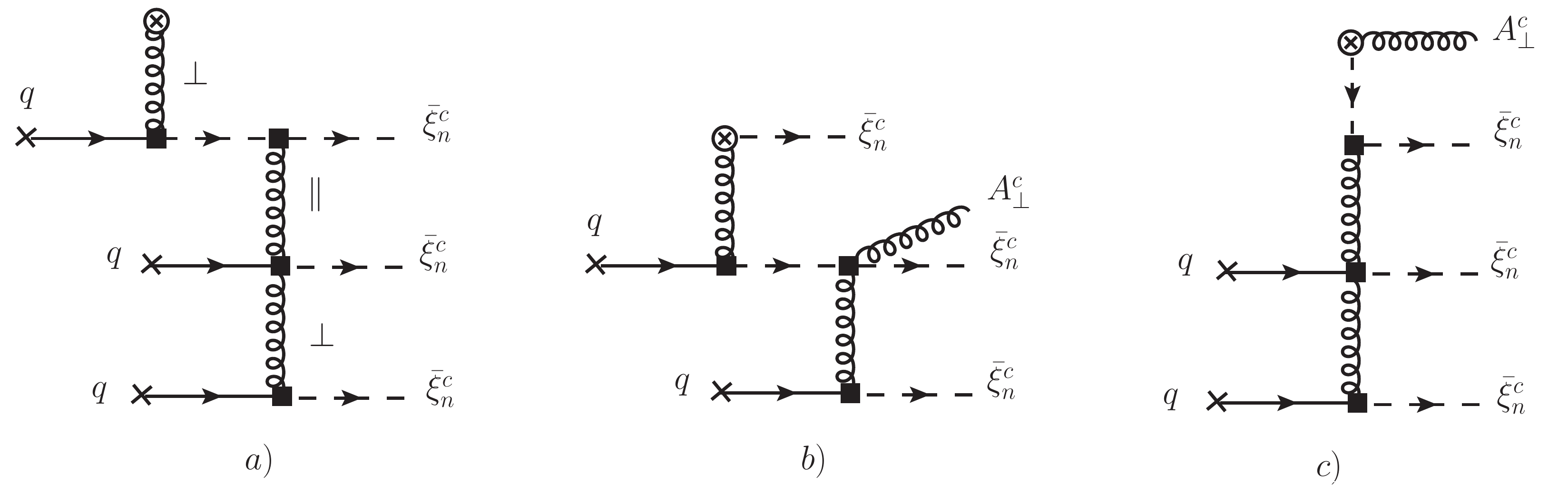}%
\caption{The SCET diagram illustrating the $T$-product in Eq.(\ref{TAbot}). The black squares show the interaction vertices, dashed quark lines denote the hard-collinear and external collinear particles. The parallel ($\|$) and transverse ($\bot$) signs  show the contractions of the appropriate hard-collinear gluon fields.}%
\label{Abot}%
\end{figure}

The simplest  operator  with the required structure and can be built from the gluon fields and is of order $\lambda^{2}$ 
\begin{equation}
O_{\mu\nu}^{(2)}=\mathcal{A}_{\bot\mu}^{n}\mathcal{A}_{\bot\nu}^{\bar{n}
}+\mathcal{A}_{\bot\nu}^{n}\mathcal{A}_{\bot\mu}^{\bar{n}}.
\label{O2g}
\end{equation}
The SCET matrix element of this operator  can be written as  
\begin{equation}
\left\langle p^{\prime}\right\vert \mathcal{A}_{\bot\mu}^{n}\mathcal{A}%
_{\bot\nu}^{\bar{n}}\left\vert p\right\rangle _{\text{{\footnotesize SCET}}%
}=g_{\alpha\beta}^{\bot}~\bar{N}_{n}1N_{\bar{n}}~\mathcal{F}^{g}%
(t)+\epsilon_{\alpha\beta}^{\bot}~\bar{N}_{n}i\gamma_{5}N_{\bar{n}%
}~\mathcal{\tilde{F}}^{g}(t),
\end{equation}
with
\bea
g_{\alpha\beta}^{\bot}=g_{\alpha\beta}-\frac12(n^{\alpha}\bar n^{\beta}+\bar n^{\alpha} n^{\beta}), \\
 \epsilon_{\alpha\beta}^{\bot}= \frac12 \epsilon_{\alpha\beta\rho\sigma}n^{\rho}\bar n^{\sigma}.
\eea
In SCET-II,  the contribution of each collinear sector yields a soft-collinear
operator at least of order $\lambda^{7}$
\begin{equation}
T\left\{  \mathcal{A}_{\bot\mu}^{n},\mathcal{L}^{(1,n)}[\bar{\xi}A_{\bot}q],
\mathcal{L}^{(1,n)}[\bar{\xi^{c}}(n\cdot A)\xi],
\mathcal{L}^{(2,n)}[\bar{\xi}^{c}(\bar{n}\cdot A)A_{\bot}q],
\mathcal{L}^{(2,n)}_ {\text{int}} [\bar{\xi}^{c}A_{\bot}q]\right\}
\sim O_{n}^{(6)}\ast J_{n}\ast {q}{q}{q},
\label{TAbot}
\end{equation}
where $J_{n}$ is the hard-collinear kernel (jet function) and the asterisks denotes the appropriate convolutions.  
The SCET interactions $\mathcal{L}^{(i,n)}_ {\text{int}}$ are shown schematically\footnote{The explicit expressions can be found in Ref.\cite{Kivel:2013sya}}. The $T$-product  in  Eq.(\ref{TAbot})  can be  illustrated with the help of the Feynman  diagrams in Fig.\ref{Abot}.  A similar $T$-product also describes the second collinear sector.
Notice that  the collinear operators  in this case are the leading-order operators. Nevertheless,  the helicity-flip  structure of the amplitude  is provided by  the chiral-odd  three-quark soft  correlator.  The total contribution associated with the operator (\ref{O2g}) is of order $\lambda^{14}$ as  required. 
However  the hard coefficient function of the gluon operator (\ref{O2g})  is subleading in $\alpha_{s}$.  In our  further analysis, we restrict our consideration to the leading-order accuracy in the hard coupling  $\alpha_{s}$. Therefore we 
neglect  the contribution of the pure gluonic operator (\ref{O2g}). 

The other  suitable operators $\mathcal{O}_{I}$ are  of order $\lambda^{3}$
and can be  built from the quark-gluon  combinations $\bar{\chi}_{n}(0)\gamma_{\bot
}^{\alpha}{\cal A}_{\bot\beta}^{n}(\lambda\bar{n})\chi_{\bar{n}}(0)$ and $\bar{\chi
}_{n}(0)\gamma_{\bot}^{\alpha}{\cal A}_{\bot\beta}^{\bar{n}}(\lambda n)\chi_{\bar{n}
}(0)$.  We find  the following two relevant scalar operators
\begin{equation}
O_{q}^{(3)}(\lambda)=~\bar{\chi}_{n}^{q}\left\{  ~\Dslash{ {\cal A}} _{\bot}^{n}+\Dslash{ {\cal A}} _{\bot}^{\bar{n}}\right\}  
\chi_{\bar{n}}^{q}+\chi_{\bar{n}}^{q}\left(  \Dslash{ {\cal A}} _{\bot}^{\bar{n}}+\Dslash{ {\cal A}} _{\bot}
^{n}\right)  \chi_{n}^{q},
\label{Oq3}
\end{equation}
\begin{equation}
\tilde{O}_{q}^{(3)}(\lambda)=\bar{\chi}_{n}^{q}\left\{
\Dslash{ \tilde{{\mathcal A}}}_{\bot}^{n}-\Dslash{ \tilde{{\cal A}}}_{\bot}^{\bar{n}}\right\}  \chi_{\bar{n}}^{q}+
\chi_{\bar{n}}^{q}\left(  \Dslash{\tilde{ {\cal A}}}_{\bot}^{\bar{n}}-\Dslash{\tilde{ {\cal A}}}_{\bot}
^{n}\right)  \chi_{n}^{q},
\label{tOq3}
\end{equation}
where the index $q$ denotes the quark flavor  and 
\begin{equation}
\tilde{\cal{A}}_{\bot\alpha}^{n}=\epsilon^{\alpha\beta}_{\bot}\mathcal{A}_{\bot\beta}^{n}.
\end{equation}

The higher order  subleading operators of this type can be constructed adding the  gluon fields $A_{\bot
}\sim\lambda$ or  $\left(  A^{n}\cdot n\right)\sim\lambda^{2}$.  Such operators will be suppressed as
$\mathcal{O}(\lambda^{5})$.  We find  that  in SCET-II these operators provide the power suppressed
contributions $\sim \mathcal{O}(\lambda^{16})$ and therefore can be neglected. 
We shall not provide a proof of  this statement in the present work  and accept  it as a plausible working assumption.  Then at
leading-order in the hard coupling $\alpha_{s}$  the power suppressed helicity-flip contribution
is only described by the two operators $O_{q}^{(3)}$ and $\tilde{O}_{q}^{(3)}$.  

In order to show  the relevance of the SCET-I operators  let us demonstrate  the mixing  of
the soft-spectator contributions  described by the operators (\ref{Oq3}) and (\ref{tOq3}) with  the hard-spectator configuration.   
Such mixing is provided by the appropriate
hard-collinear $T$-products  which describe the matching  on the SCET-II soft-collinear operators.   In order to simplify this discussion we consider 
the contractions of the hard-collinear fields in each hard-collinear sector  separately (the collinear and soft fields are considered as external)
\bea
T\{O_{q}^{(3)}\}&=&T\{\bar{\chi}_{n} \Dslash{ {\cal A} } _{\bot}^{n}\}\  T\{\chi_{\bar{n}}\}.
\label{TO=TnTbn}
\eea  
The total soft-collinear operator  is given by the suitable soft-collinear combinations from  each hard-collinear sector.  
The $T$-product of the hard-collinear   field  $\chi_{n,\bar{n}}$   can be interpreted  as a transition of the  hard-collinear quark and two soft spectator quarks  into
three collinear quarks or vice versa, schematically 
\bea
T\{\chi_{\bar{n}}\}\simeq \bar q\bar q*J_{\bar n}*O_{\bar n}^{(6)}\sim \lambda^{6}.
\eea
A combination of such  $T$-products  yields the soft-collinear operator describing  the soft-spectator contribution  for the  leading  amplitudes $T_{2,4,6}$, see details in Ref.\cite{Kivel:2013sya}.  The  configurations with the subleading collinear operators  can be generated  from the hard-collinear sub-operator $\bar{\chi}_{n} \Dslash{ {\cal A} } _{\bot}^{n}$ in Eq.(\ref{TO=TnTbn}).  
For instance,  matching on a twist-4 collinear operator $O_{n}^{(8)}\sim \bar \xi^{c}_{n}\bar \xi^{c}_{n}\bar \xi^{c}_{n}A^{n}_{\bot c}$  can be described by the  following  $T$-products
\bea
T\left\{ 
\bar{\chi}_{n}^{c} \Dslash{ {\cal A} } _{\bot}^{n},  
\mathcal{L}^{(1,n)}_{\text{int}} [\bar \xi A_{\bot}q],
\mathcal{L}^{(2,n)}_{\text{int}} [\bar \xi^{c} A^{c}_{\bot}A_{\bot} \xi ] ,
\mathcal{L}^{(2,n)}_{\text{int}}[\bar \xi^{c}A_{\bot}q] 
\right\} 
\sim O_{n}^{(8)}*J_{n}*qq\sim\lambda^{8},
\label{TL4L2}
\eea
\bea
T\left\{ 
\bar{\chi}_{n} \Dslash{ \mathcal{A} }^{n} _{\bot c},  \mathcal{L}^{(1,n)}_{\text{int}} [\bar \xi^{c}(n\cdot A)\xi], \mathcal{L}^{(2,n)}_{\text{int}}[\bar \xi^{c}(\bar n\cdot A)A_{\bot}q],
\mathcal{L}^{(2,n)}_{\text{int}}[\bar \xi^{c}A_{\bot}q] 
\right\} 
\sim O_{n}^{(8)}*J_{n}*qq\sim\lambda^{8}.
\label{TL1L2-2}
\eea
The diagrams described by these $T$-products are shown in Fig.\ref{Abot}~$b)$ and $c)$, respectively.  We also accept that the collinear  fields which appear  in the SCET-I operators in Eqs.(\ref{TL4L2}) and (\ref{TL1L2-2}) are generated by the substitution $\phi_{hc}\rightarrow \phi_{hc}+\phi_{c}$ performing matching onto SCET-II operators.  Combining results of the two  hard-collinear $T$-products one obtains  a soft-collinear operator
\bea
T\{O_{q}^{(3)}\}\simeq  O_{n}^{(8)}*J_{n}*qq\bar q\bar q*J_{\bar n}*O_{\bar n}^{(6)}\sim \lambda^{14},
\eea
which consist of  the same collinear operators  as the appropriate hard-spectator contribution $O_{n}^{(8)}*\tilde{T}*O_{\bar n}^{(6)}$.  
Here we will not study  the  structure of  all possible collinear  contributions.  We expect that the two presented examples clearly illustrate the 
presence of the soft-spectator contributions in Eq.(\ref{TJJ}).   In the following discussion  we  
assume that at the leading-order in $\alpha_{s}$ the soft-spectator contribution 
is only described by the matrix elements of the two operators (\ref{Oq3}) and (\ref{tOq3}).   

Let us consider SCET matrix elements of these operators.  They can be described as 
\begin{equation}
\left\langle p^{\prime}\right\vert \sum_{q=u,d} e^{2}_{q}\ O_{q}^{(3)}(\lambda)\left\vert p\right\rangle
_{\text{{\footnotesize SCET}}}=m~\bar{N}_{n}1N_{\bar{n}}\int_{0}^{1}%
d\tau\left\{  ~e^{i\lambda(p^{\prime}\cdot\bar{n})\tau}~+e^{-i\lambda(p\cdot n)\tau
}\right\}    \mathcal{G}(\tau,t), 
\label{O3me}
\end{equation}
\begin{equation}
\left\langle p^{\prime}\right\vert  \sum_{q=u,d} e^{2}_{q}\ \tilde{O}_{q}^{(3)}(\lambda)\left\vert
p\right\rangle _{\text{{\footnotesize SCET}}}=m~\bar{N}_{n}\gamma_{5}%
N_{\bar{n}}\int_{0}^{1}d\tau\left\{  ~e^{i\lambda(p^{\prime}\cdot\bar{n})\tau
}~+e^{-i\lambda(p\cdot n)\tau}\right\}  \tilde{\mathcal{G}}(\tau,t)  ,
\label{tO3me}
\end{equation}
where on the {\it lhs} we defined the required flavor combinations. Dimensionless amplitudes $ \mathcal{G}$ and $\tilde{\mathcal{G}}$ also depend
on the factorization scale $\mu_{F}$ which  is not shown for simplicity. This scale 
separates contributions from  the  hard and  hard-collinear regions. The SCET-I amplitudes 
describes the dynamics associated with  hard-collinear scale $\sim\sqrt{\Lambda Q}$  and  soft scale $\sim \Lambda$.  Therefore these 
amplitudes are   functions of the momentum transfer. The 
fraction $\tau$ can be interpreted as the fraction of the collinear momentum
carried by the hard-collinear transverse gluon.  

In order to obtain a formal  factorization formula for the amplitudes $T_{1,3,5}$  one  has to take the matrix element from Eq.(\ref{TJJ}) 
and use  for the soft-spectator  contributions on the {\it rhs} the matrix elements defined in Eqs.(\ref{defF1}, \ref{tO3me}) and (\ref{O3me}).
On the other hand, the nucleon spinors in the parametrization (\ref{defM}) appearing on the {\it lhs}  must be rewritten in terms of the large components 
defined in (\ref{Nn}).   

For illustration let us consider  the calculation of  amplitudes $T_{1,2}$.  These amplitudes can be easily singled out  using the contraction 
\bea
-\mathcal{T}_{12}^{\mu\nu}\bar{N}(p^{\prime})\mathcal{A}^{\mu\nu} N(p)=\bar{N}(p^{\prime})\left(  T_{1}+\Dsl[K]{7}~T_{2}\right)N(p)=-\mathcal{T}_{12}^{\mu\nu}  \langle p' |  \sum\tilde{C}^{\mu\nu}\ast {\mathcal O}_{I}+\sum O_{n}^{(i)}\ast\tilde
{T}^{\mu\nu}\ast O_{\bar{n}}^{(j)} | p\rangle.
\label{T13ope}
\eea 
The {\it lhs}  can be rewritten as  
\bea
\bar{N}(p^{\prime})\left(  T_{1}+\Dsl[K]{7}~T_{2}\right)N(p)
=\bar{N}_{n}1 N_{\bar n}(1+\kappa) \left\{  T_{1}+\frac{m(s-u)}{4m^{2}-t}~T_{2}\right\} +\bar{N}_{n}\Dslash{q}_{\perp}N_{\bar n} \left(1-\kappa \right)  T_{2},
\label{lhsT13ope}
\eea
where we used 
\bea
\bar{N}(p^{\prime})\Dsl[K]{7}N(p)&=&\bar{N}_{n}\Dslash{q}_{\perp}N_{\bar n}\left(1-\kappa \right)+\bar{N}_{n}1 N_{\bar n}\frac{m}{W}K\cdot(n+\bar n),
\\
\bar{N}(p^{\prime})1 N(p)&=&\bar{N}_{n}1 N_{\bar n}(1+\kappa).
\eea  
The {\it rhs}  of  (\ref{T13ope})  can be written  as 
\bea
-\mathcal{T}_{12}^{\mu\nu}  \langle p' |  \sum\tilde{C}^{\mu\nu}\ast {\mathcal O}_{I}+\sum O_{n}^{(i)}\ast\tilde
{T}^{\mu\nu}\ast O_{\bar{n}}^{(j)} | p\rangle
=
\bar{N}_{n}\Dslash{q}_{\perp}N_{\bar n}\left\{ C_{2}(s,t) \mathcal{F}_{1}+ {\Psi_{\text{tw3}}}\ast H_{2}(s,t)\ast\Psi_{\text{tw3}}   \right\}
\\
+\bar{N}_{n}1 N_{\bar n}
\left\{ m~C_{1}(s,t)\ast~\mathcal{G}(t)+{\Psi_{\text{tw3}}}\ast H_{1}(s,t)\ast\Psi_{\text{tw4}} \right\} +\ldots
\label{T13operhs}
\eea 
Here  $C_{1,2}$ and $H_{1,2}$ denote  the  momentum space hard coefficient functions  in the soft- and hard-spectator contributions, respectively. The asterisks denote the convolution integrals with respect to the collinear fractions,  the hard-spectator contributions are shown schematically, $\Psi_{\text{tw3}}$, $\Psi_{\text{tw4}}$ denote the nucleon distribution amplitudes of twist-3 and twist-4, respectively.   

Comparing  Eqs.(\ref{lhsT13ope}) and (\ref{T13operhs})  one obtains
\begin{equation}
T_{2}\simeq (  1-\kappa)  ^{-1}\left\{  ~C_{2}(s,t)\mathcal{F}(t)+\Psi_{\text{tw3}}\ast H_{2}(s,t)\ast\Psi_{\text{tw3}} \right\} ,
\label{T2}
\end{equation}%
\begin{equation}
T_{1}\simeq-\frac{m(s-u)}{4m^{2}-t}~T_{2}+\left(  1+\kappa \right)
^{-1}\left\{  m~C_{1}(s,t)\ast\mathcal{G}(t)+\Psi_{\text{tw3}}\ast H_{1}(s,t)\ast\Psi_{\text{tw4}}\right\}.
 \label{T1full}
\end{equation}
Using Eq.(\ref{defT1b}) one also finds
\begin{equation}
\bar T_{1}\simeq \left(  1+\kappa \right)
^{-1}\left\{  m~C_{1}(s,t)\ast\mathcal{G}(t)+\Psi_{\text{tw3}}\ast H_{1}(s,t)\ast\Psi_{\text{tw4}}\right\}.
 \label{T1b}
\end{equation}
This clarify the  substitution introduced in Eq.(\ref{defT1b}): such redefinition removes the kinematical  part  associated with $T_{2}$ from the expression for $T_{1}$ in Eq.(\ref{T1full}).  The soft-spectator contribution of the  amplitude $\bar T_{1}$   is only defined by the  subleading SCET amplitude $\mathcal {G}(\tau, t)$.  We also keep the power suppressed factors $(1\pm\kappa)$ in  Eqs.(\ref{T2}-\ref{T1b})  as the  kinematical power corrections. 

\begin{figure}[ptb]%
\centering
\includegraphics[
height=0.6899in,
width=3.6538in
]%
{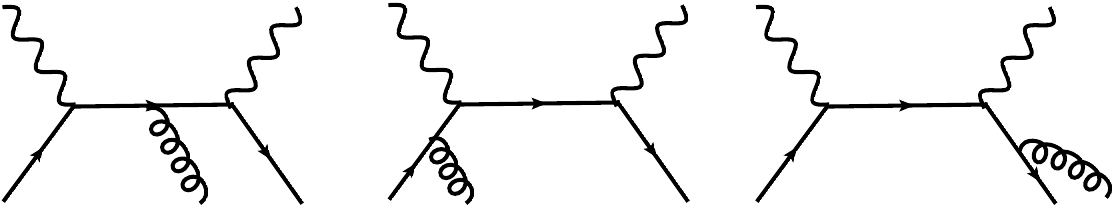}%
\caption{The tree diagrams required for the matching onto subleading
operators. }%
\label{sub-o-diagrams}%
\end{figure}

The similar  calculations  give 
\begin{equation}
\bar{T}_{3}=\left(  1+\kappa \right)  ^{-1}\left\{  mC_{3}
(s,t)\ast\mathcal{G}(t)+\Psi_{\text{tw3}}\ast H_{3}(s,t)\ast\Psi_{\text{tw4}}\right\}  ,
\end{equation}
\begin{equation}
T_{4}=\left(  1-\kappa \right)  ^{-1}\left\{  ~C_{4}(s,t)\mathcal{F}(t)+\Psi_{\text{tw3}}\ast H_{4}(s,t)\ast\Psi_{\text{tw3}}\right\}  ,
\end{equation}
\begin{equation}
T_{5}=\left(  1-\kappa \right)^{-1} \left\{  mC_{5}(s,t)\ast\mathcal{\tilde{G}}(t)+\Psi_{\text{tw3}}\ast H_{5}(s,t)\ast\Psi_{\text{tw4}}\right\}  ,
\label{T5}
\end{equation}
\begin{equation}
T_{6}=\left(  1+\kappa \right)  ^{-1}\left\{  ~C_{6}%
(s,t)\mathcal{F}(t)+\Psi_{\text{tw3}}\ast H_{6}(s,t)\ast\Psi_{\text{tw3}} \right\}  .
\label{T6}
\end{equation}
The hard coefficient functions $C_{2,4,6}$ can be found in Ref.\cite{Kivel:2013sya}.  The subleading  coefficient functions $C_{1,3,5}$ can be computed from the diagrams in Fig.\ref{sub-o-diagrams} 
and read
\begin{equation}
C_{1}(s, t,\tau)=-\frac{1}{1-\tau}
\frac{\hat t} { \hat s\hat u}+2\left(  \frac{\hat t}{\hat s\hat u}+\frac{1}{\hat t}\right)  ,
\end{equation}%
\begin{equation}
C_{3}( s,t, \tau)=-\frac
{1}{1-\tau}\frac{\hat t}{\hat s\hat u}-\frac{2}{\hat t},
\end{equation}%
\begin{equation}
C_{5}( s, t,\tau)=\frac{\tau}{1-\tau}\frac{\hat t}{\hat s\hat u},
\end{equation}
where $\tau$ is  the gluon fraction, $0<\tau<1$ and the hat denotes  the partonic (massless) Mandelstam variables related to the scattering angle in c.m.s.  as:
\begin{equation}
\hat t=-\frac{\hat s}{2}(1-\cos\theta),~\hat u=-\frac{\hat s}{2}(1+\cos\theta),~\ \hat s+\hat t+\hat u=0.
\end{equation}

To calculate the observables of Eq.(\ref{W12}) and (\ref{W22})   the following combinations are needed
 \begin{equation}
\bar{T}_{1}-\bar{T}_{3}=\left(  1+\kappa \right)^{-1}
\left\{
m~\Delta (s,t)\int_{0}^{1}d\tau~\mathcal{G}(\tau,t)+\Psi_{\text{tw3}}\ast\left(
H_{1}-H_{3}\right)  \ast\Psi_{\text{tw4}}\right\}  ,
\label{DT13}
\end{equation}
\begin{align}
\bar{T}_{1}+\bar{T}_{3}  &  =\left(  1+\kappa \right)  ^{-1}\left\{ m\Sigma(s,t) ~\int_{0}^{1}d\tau~\frac{\tau}{1-\tau}\mathcal{G}%
(\tau,t)~+\Psi_{\text{tw3}}\ast {\left(  H_{1}-H_{3}\right)  }\ast\Psi_{\text{tw4}}\right\}, 
\label{ST13}
\end{align}
where
\begin{equation}
\Delta (s,t)=2 \frac{\hat t}{\hat s\hat u} \left( 1+2\frac{\hat s\hat u}{\hat t^{2}}\right),\   \Sigma(s,t)=-2\frac{\hat t}{\hat s\hat u}.
\end{equation}
The soft- and hard-spectator contributions in the expressions for the amplitudes $T_{i}$ (\ref{T2})-(\ref{T6})   have  endpoint singularities which  cancel in their  sum.

\section{Phenomenology}
\label{phenom}
The estimates based on the hard-spectator scattering mechanism predict an order of magnitude smaller cross section for the WACS cross section, see \textit{e.g.}  
Refs.\cite{Vanderhaeghen:1997my, Brooks:2000nb, Thomson:2006ny}.
Therefore  we assume that the soft-spectator contributions  dominate over  the hard-spectator ones in the relevant kinematical region. 
 It is convenient to introduce the function $\mathcal{R}(s,t)$ as:
\begin{equation}
T_{2}=C_{2}(s,t)\left(  1-\kappa\right)  ^{-1}\left\{
~\mathcal{F}(t)+\Psi\ast\frac{H_{2}(s,t)}{C_{2}(s,t)}\ast
\Psi\right\}  \equiv C_{2}(s,t)\mathcal{R}(s,t).
\label{defR}
\end{equation}  
In the kinematical region where the soft-spectator contribution dominates,  the introduced  ratio $\mathcal{R}(s,t)$  must  be almost $s$-independent, i.e.
\begin{equation}
\mathcal{R}(s,t)\simeq \mathcal{R}(t),
\label{Rst}
\end{equation}
because the $s$-dependent term  in Eq.(\ref{defR}) is only given by the hard-spectator contribution.  The  expressions for the  other helicity-conserving amplitudes can also be defined in  terms of this ratio up to small next-to-next-to-leading order corrections \cite{Kivel:2013sya}
\begin{equation}
T_{4}\simeq C_{4}(s,t)\mathcal{R}(t)+\mathcal{O}(\alpha_{s}^{2}),
\label{T4=C4R}
\end{equation}%
\begin{equation}
T_{6}\simeq\sqrt{\frac{-t}{4m^{2}-t}}~C_{6}(s,t)\mathcal{R}(t)+\mathcal{O}(\alpha_{s}^{2}), 
\label{T6=C6R}%
\end{equation}
The  similar expressions for the amplitudes $T_{2,4,6}$ have already  been considered in the Refs. \cite{Kivel:2012vs,Kivel:2013sya} but  without  the power suppressed factor $\sqrt{{-t}/(4m^{2}-t)}$ in Eq.(\ref{T6=C6R}). This factor is  part of the full kinematical power correction which was neglected in the previous work. 

\begin{figure}[ptb]
\centering
\includegraphics[width=3.0in]{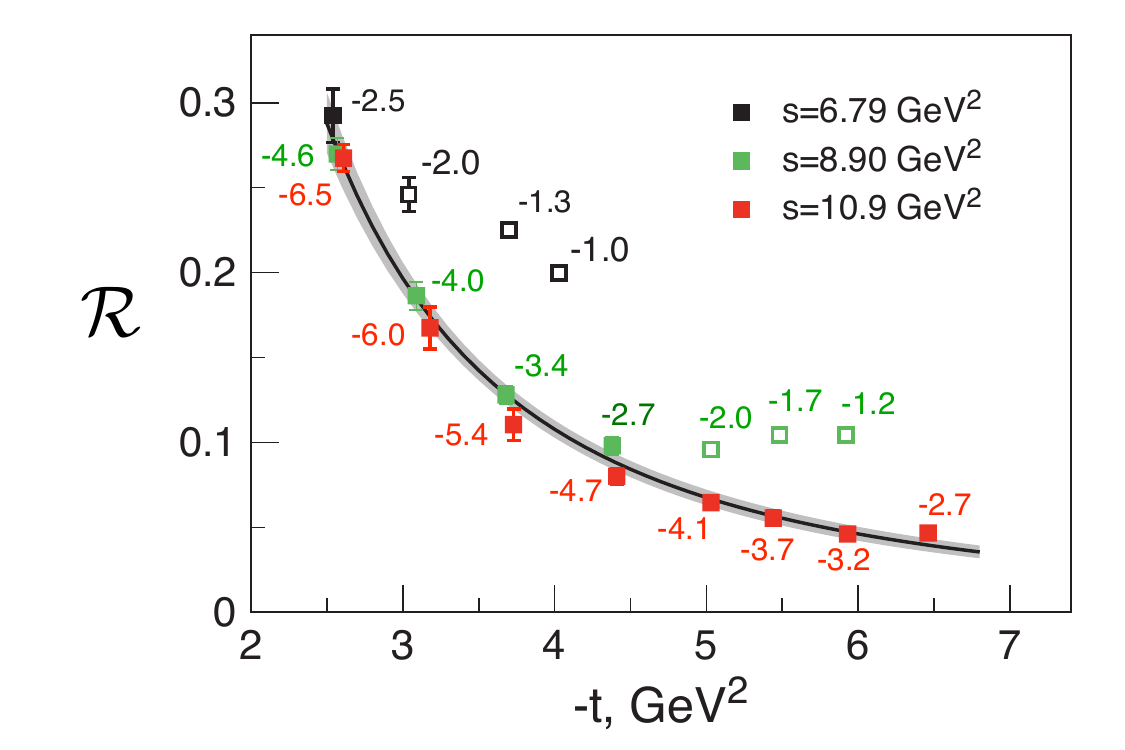}
\includegraphics[width=3.0in]{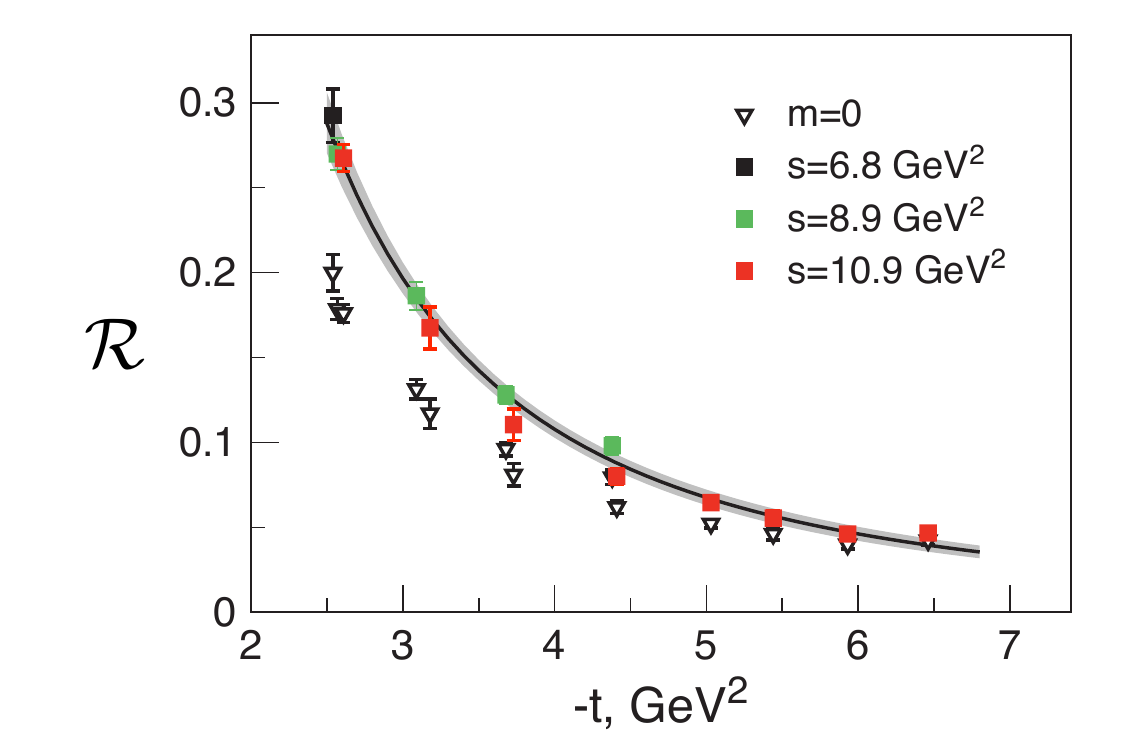}
\caption{{\bf Left:} The ratio $\mathcal{R}$ as function of $t$. The corresponding  values of $u$  are shown by the numbers next to the symbols.   The
open squares mark out  the points with $-u<2.5$GeV$^{2}$. The solid line represents the empirical fit of Eq.(\ref{Rfit}), the gray bands show the $99\%$ confidence interval.    {\bf Right:} The ratio $\mathcal{R}$ as function of $t$, where only data with $-u>2.5$GeV$^{2}$ are shown. 
 The open triangles  show the values of $\mathcal{R}$ obtained  without the  kinematical power corrections, i.e. by setting $m=0$.   }
 \label{RLOkinadd}
\end{figure}

Deriving  the formulae (\ref{T4=C4R}) and (\ref{T6=C6R})  we  use  that all three amplitudes $T_{2,4,6}$ depend on the same $t$-dependent SCET amplitude $\mathcal{F}(t)$  and factorize  multiplicatively.   For the helicity flip amplitudes the situation is more complicated because in
this case one deals with  the convolution integrals  of the hard coefficient functions with  two different SCET amplitudes. This leads to  a more complicated  structure of   power suppressed contributions.  In order to proceed further, we introduce the following  three  amplitudes $G_{0}(s,t)$, $G_{1}(s,t)$ and $\tilde G_{1}(s,t)$
\begin{align}
G_{0}(s,t) =(  1-\kappa) ^{-1}\left\{  ~\int_{0}^{1}d\tau~\mathcal{G}(\tau,t)~+\Psi_{\text{tw3}}\ast\frac{\left(  H_{1}-H_{3}\right)(s,t)  }
{m \Delta(s,t) }\ast\Psi_{\text{tw4}}\right\},
\label{defG0}
\end{align}
\begin{align}
G_{1}(s,t)  =\left(  1-\kappa \right)  ^{-1}\left\{  ~\int_{0}^{1}d\tau~\frac{\tau}{1-\tau}\mathcal{G}
(\tau,t)~+\Psi_{\text{tw3}}\ast\frac{\left(  H_{1}+H_{3}\right)(s,t)  }{m\Sigma(s,t)}\ast\Psi_{\text{tw4}}\right\},
\label{defG1}
\end{align}
and 
\begin{align}
\tilde{G}_{1}(s,t)= \left(  1-\kappa \right)  ^{-1}\left\{
\int_{0}^{1}d\tau\frac{\tau}{1-\tau}\mathcal{\tilde{G}}(\tau, t)+\Psi_{\text{tw3}}
\ast\frac{H_{5}(s,t)}{C_{5}(s,t)}\ast\Psi_{\text{tw4}}\right\} .
\label{deftG1}
\end{align}
Analogously to $\mathcal{R}(s,t)$   these new functions are defined using the expressions for the amplitudes obtained in Eqs.(\ref{T5}),(\ref{DT13}) and (\ref{ST13}).
Assuming   the dominance of the soft-spectator part we again can expect  that  the $s$-dependence of these functions is  weak 
\begin{equation}
 G_{0,1}(s,t)\simeq G_{0,1}(t),~\ \tilde {G}_{1}(s,t)\simeq\tilde{G}_{1}(t).
\end{equation}
Under such assumption,  we obtain 
\begin{align}
\bar{T}_{1}-\bar{T}_{3} \simeq  m\Delta(s,t)\sqrt{\frac{-t}{4m^{2}-t}}~G_{0}(t), \label{T1-T3}
\end{align}
\begin{align}
\bar{T}_{1}+\bar{T}_{3}\simeq  m\Sigma(s,t)\sqrt{\frac{-t}{4m^{2}-t}}~G_{1}(t), \label{T13=SG1}
\end{align}
\begin{align}
{T}_{5} \simeq   m~C_{5}(s,t)\tilde{G}_{1}(t). \label{T5=C5G1}
\end{align}
Substituting  the obtained  expressions for the amplitudes $T_{i}$ in  Eq.(\ref{W00})  for $W_{00}$ we obtain
\begin{align}
W_{00}  &  \simeq\frac{m^{4}-su}{4m^{2}-t}(-t)\left\{  \frac{1}{2}
 (C_{2} ^{2}+ C_{4} ^{2})
+C_{6} ^{2}\right\}  \mathcal{R}^{2}(t)
 +m^{2}(-t)\left\{  ~\Delta^{2} G_{0}^{2}%
(t)+\Sigma^{2}G_{1}^{2}(t) + C^{2}_{5} \tilde{G}_{1}^{2}(t)
 \right\}.
 \label{W00R}
\end{align}
The {\it rhs} of Eq.(\ref{W00R})  depends  on  four unknown $t$-dependent functions $\mathcal{R}$,  $G_{0,1}$ and $\tilde G_{1}$.  Three
of these functions are related to the helicity-flip amplitudes.  One can expect that at large $-t$ these functions are  smaller than  $\mathcal{R}$.
For instance, for the case of the nucleon form factors, data at large momentum transfer show that  $G_{E}/G_{M}\ll 1$. 
Let us also assume that  the helicity-flip amplitudes $G_{0,1}$ and $\tilde G_{1}$  in WACS are  also  smaller than $\mathcal{R}$.  
This assumption  is also plausible because  the amplitudes  $G_{0,1}$  
are defined by the similar subleading operators as  the form factor $G_{E}$ within the SCET formalism, see, {\it e.g.}  \cite{Kivel:2010ns}.     
Neglecting the helicity-flip contributions  in Eq.(\ref{W00R}) ($G_{0}\approx G_{1}\approx \tilde{G}_{1}=0$) one can use the 
cross section data in order to extract the ratio $\mathcal{R}$ and to check
the scaling  behavior implied by Eq.(\ref{Rst}). We recall, that the leading-order coefficient
functions $C_{2,4,6}$ read \cite{Kivel:2013sya}
\begin{equation}
C_{2}=-C_{4}=\frac{\hat{s}-\hat{u}}{\hat{s}\hat{u}}=-\frac{1}{s}\frac
{3+\cos\theta}{1+\cos\theta}, \ \
C_{6}=\frac{\hat{t}}{\hat{s}\hat{u}}=\frac{1}{s}\frac{1-\cos\theta}%
{1+\cos\theta}. \label{CiLO}%
\end{equation}
For the scattering angle $\theta$ in Eqs.(\ref{CiLO})  we use the substitution
\begin{equation}
\cos\theta=1+\frac{2st}{(s-m^{2})^{2}},
\end{equation}
which also includes the power suppressed  terms which are considered  as a part of the kinematical corrections.  
The obtained  results for $\mathcal{R}$ are shown  in Fig.\ref{RLOkinadd}.  

The left plot in Fig.\ref{RLOkinadd} shows the value of  $\mathcal{R}$ as a function of the momentum transfer for  $-t\geq 2.5$~GeV$^{2}$.  As it was  assumed above, see Eq.(\ref{Rst}), the extracted values of  $\mathcal{R}$  are expected to show only  a small sensitivity to  $s$ when the soft spectator mechanism dominates.  From  Fig.\ref{RLOkinadd}  we see that this approximate scaling behavior is observed  in the region where $-u\geq 2.5$~GeV$^{2}$.  Hence we can adopt this value as a phenomenological lower limit of applicability of the  described approach. 
 For smaller values of $u$ the  extracted  values of  $\mathcal{R}$  (shown by the open squares) demonstrate already a clear sensitivity to $s$.  Thus one can observe  that  for $-u=1.3$~GeV$^{2}$ ($-t=3.7$~GeV$^{2}$) the obtained value of $\mathcal{R}$ is about a factor 2 larger than  the scaling curve.  This observation clearly demonstrates that  the given  approach  can not  describe  the  cross section data  at  small values of $u$.  
 
 The solid line in both plots in Fig.\ref{RLOkinadd}  corresponds to the fit of the  points with $-t,-u \geq 2.5$~GeV$^{2}$  by a simple empirical  ansatz
\bea
\mathcal{R}(t)=\left( \frac{\Lambda^{2}}{-t}\right)^{\alpha},
\label{Rfit}
\eea
where $\Lambda$ and $\alpha$ are free parameters.  For their values we obtain  $\Lambda=1.17\pm0.01$~GeV and $\alpha=2.09\pm 0.06$. The shaded area in Fig.\ref{RLOkinadd}  shows the  confidence interval with CL$=99\%$. 
 
 On the right plot in  Fig.\ref{RLOkinadd}  we show the effect of the  kinematical power suppressed contributions.  The empty triangles show the values of  $\mathcal{R}$  obtained   without  kinematical power corrections  with  $m=0$.  The difference  between the values of $\mathcal{R}$ extracted with and without power suppressed contributions is about $30\%$ at the lower value $-t\approx 2.5$~GeV$^{2}$.    Let us notice that   the  values of $\mathcal{R}$ obtained in this work are somewhat larger than  ones obtained in Refs.\cite{Kivel:2012vs, Kivel:2013sya}. This difference is explained by the incomplete  description of the  kinematical power corrections in the previous works.  
   
\begin{figure}[ptb]
\centering
\includegraphics[height=2.0955in]{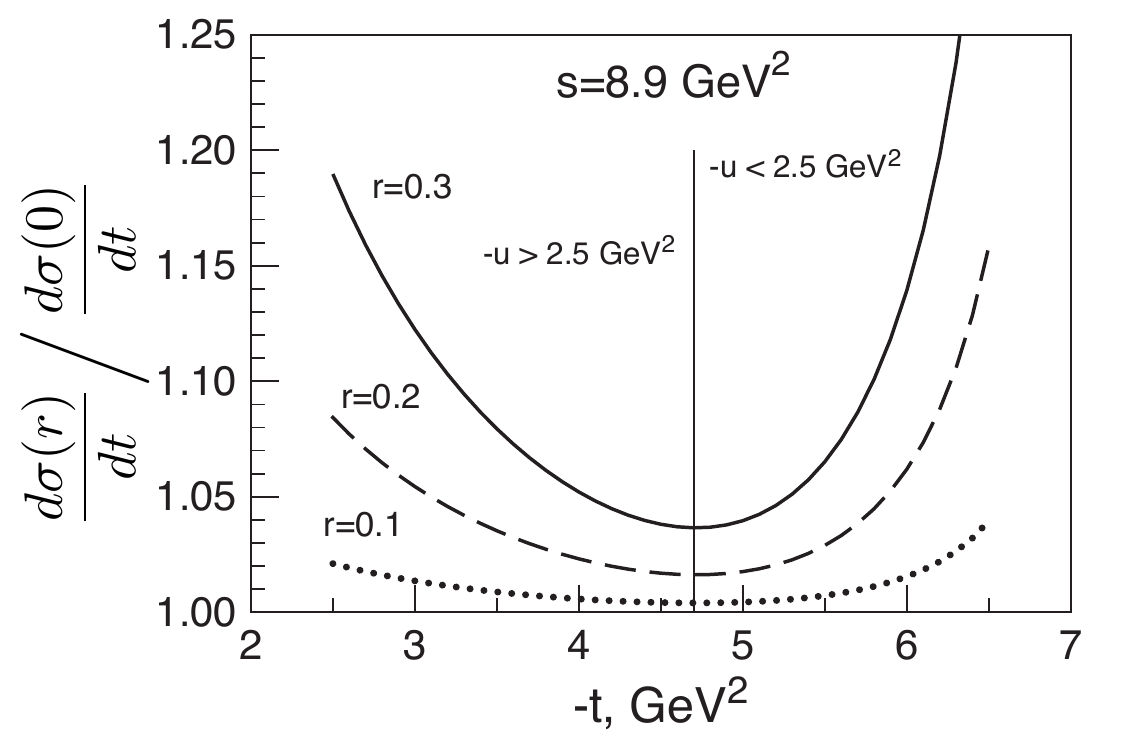}
\caption{The ratio of
the cross sections as a function of momentum transfer at different values of $r_{0}=r_{1}=\tilde r_{1}=r$. }
\label{hflip}
\end{figure}

The consistent  results  for  the ratio  $\mathcal{R}$, extracted in the present framework,  indicate that the assumption about the relative smallness of  helicity-flip amplitudes is probably correct.  We next investigate if  one can obtain an estimate  of the helicity flip amplitudes from the cross section data.  For this purpose,  it is convenient to introduce the following ratios
\begin{equation}
\frac{G_{0}(t)}{\mathcal{R}(t)}=r_{0}(t),~\frac{G_{1}(t)}{\mathcal{R}%
(t)}=r_{1}(t),~~\ ~\frac{\tilde{G}_{1}(t)}{\mathcal{R}(t)}=\tilde{r}%
_{1}(t).\ \ \label{def-ri}%
\end{equation}
In the following discussion we assume that numerically these three quantities  are of the same order  and small
\bea
|r_{0}|\sim |r_{1}|\sim |\tilde r_{1}|< 1.
\eea 
In order to see the relevance of the different subleading contributions  let us consider the following  ratio  of the cross sections  at 
$s=8.9$~GeV$^{2}$ and $-t=2.5$~GeV$^{2}$, which can be expressed  as
\bea
\frac{d\sigma(r_{0},r_{1},\tilde{r}_{1})}{dt}/\frac{d\sigma(0,0,0)}{dt}=1+2.08 r_{0}^{2}+0.02 r_{1}^{2}+ 0.005 \tilde{r}_{1}^{2},
\label{sigma-r}
\eea 
 One can see that the largest  numerical impact is  provided by the contribution proportional to  $r_{0}$, the other two contributions  in Eq.(\ref{sigma-r}) have  very small coefficients and therefore their numerical impact is negligible.  This  observation also remains valid   for other values of $s$ in the region $-t,-u\geq 2.5$~GeV$^{2}$.   In Fig.\ref{hflip} we show the cross section ratio of Eq.(\ref{sigma-r}) at $s=8.9$~GeV$^{2}$ as a function of  momentum transfer. For simplicity we take the same values for  all ratios, i.e.  $r_{0}=r_{1}=\tilde r_{1}=r$.  
We can see that correction from the helicity-flip contributions are largest    at small $-t$ and  smallest  at the boundary where $-u\simeq 2.5$~GeV$^{2}$.   For illustration we also show the backward region where $-u\leq 2.5$~GeV$^{2}$  and our description is not applicable. One can see that in this region   the  contribution of the subleading amplitudes  grows and becomes more and  more important.  This can also be understood from Eq.(\ref{W00R}). The kinematical coefficient in front of $\mathcal{R}^{2}$   disappears in the backward  region because $(m^{4}-su)\rightarrow 0$.  Due to  the relative smallness   
of the  contribution proportional to  $\mathcal{R}^{2}$  in the cross section at small $-u$, the helicity-flip terms become more important.  

The relative smallness of the contributions with unknown $r_{1}$ and $\tilde r_{1}$ allows one to exclude them from the consideration and  perform an analysis of the cross section data  in order to extract the values of  $\mathcal{R}(t)$ and to constrain  $G_{0}(t)$.
   Each data point provides an inequality  $d\sigma_{min}\leq \alpha \mathcal{R}^{2}+\beta G_{0}^{2}\leq d\sigma_{max}$ where 
   $d\sigma_{max,min}=d\sigma\pm\Delta$ is the maximal and minimal  experimental values  of the cross section and $\alpha, \ \beta$ are known coefficients.  In order to find the restrictions on two unknown  quantities $\mathcal{R}^{2}$ and  $G_{0}^{2}$  one needs  at least two data points at the same $t$ and different $s$. The largest effect from  $G_{0}$  is expected  at small momentum transfer,  see Fig.\ref{hflip}.   Therefore   we consider  three data points at $-t\simeq 2.5$~GeV$^{2}$ and $s=6.8,8.9,10.9$~GeV$^{2}$  that  provide us with  three couples of  inequalities.  Combining the constraints from  each  set of  inequalities  we obtain the following restrictions:  $\mathcal{R} = 0.273-0.279$  and  $G_{0}=0.0-0.045$. The obtained value of $\mathcal{R}$ is   within the confidence interval shown in Fig.\ref{RLOkinadd}.  This results allows us  to estimate  the upper bound for the ratio $r_{0}$ 
\bea
|r_{0}(-t=2.5~\text{GeV}^{2})|\leq 0.16.
\label{r0est}
\eea  
 From Fig.\ref{hflip} it is also seen that in this case  the contribution  to  the cross section provided by  $r_{0}$ is below $10\%$.  Such uncertainty is  comparable with the   theoretical  uncertainties  such as  next-to-leading corrections or the hard-spectator corrections. Hence the result  (\ref{r0est})  must be understood as a  qualitative estimate.   

Let us  study the  effect of  subleading  amplitudes in  the asymmetries described in  Sec.{\ref{kin}}.  The  asymmetries $\KLL$ and $\KLS$ have  already  been measured at JLab in  two experiments: for large $-t$ but relatively  small $-u=1.1$~GeV$^{2}$ \cite{Hamilton:2004fq}  and in the more appropriate kinematical region for the present work \cite{E07002} (the latter analysis is not yet completed). One more experiment has  recently been suggested  in order to measure  the initial state helicity correlation  $\ALL$ in WACS  \cite{PR1214006}.   

As  we concluded above the presented  approach is not applicable in the region of small  $-u<2.5$~GeV$^{2}$.  Hence we can not use it in order to 
 describe asymmetries presented in  Ref. \cite{Hamilton:2004fq}.  Therefore despite the numerical results obtained in Ref. \cite{Kivel:2013sya},  an agreement with $\KLL $  should only be interpreted as qualitative.  
 However, the here obtained  results  can be used  for  estimates of the  asymmetries in the other experiments with  more suitable kinematics, see Table~\ref{kintab}. 
\begin{table}[h]
\caption{The kinematical regions in the two experiments of Refs.\cite{E07002,PR1214006} . }
\centering
\vspace*{0.3cm}
\begin{tabular}[c]{|c|c|c|c|}
\multicolumn{4}{c}{ $\KLL$ at  $s=9$~GeV$^{2}$, Ref.\cite{E07002} } \\
\hline
$\theta$ & $70^{o}$ & $90^{o}$ & $110^{o}$\\\hline
$-t,$~GeV$^{2}$ & $2.4$ & $3.6$ & $4.9$\\\hline
$-u,$~GeV$^{2}$ & $4.8$ & $3.5$ & $2.3$\\\hline
\end{tabular}
\hspace*{1cm}
\begin{tabular}[c]{|c|c|c|c|}
\multicolumn{4}{c}{ $\ALL$ at $s=8$~GeV$^{2}$, Ref.\cite{PR1214006} } \\
\hline
$\theta$ & $60^{o}$ & $90^{o}$ & $136^{o}$\\\hline
$-t,$~GeV$^{2}$ & $1.6$ & $3.1$ & $5.4$\\\hline
$-u,$~GeV$^{2}$ & $4.6$ & $3.0$ & $0.8$\\\hline
\end{tabular}
\label{kintab}
\end{table}

Using the leading-order expressions (\ref{defR})-(\ref{T6=C6R}) and (\ref{CiLO}) the different combinations of the amplitudes appearing in Eqs.(\ref{W12}),(\ref{W22}) can be presented as follows: 
\begin{equation}
\left(  T_{2}-T_{4}\right)  T^{*}_{6}\simeq2C_{2}C_{6}\sqrt{\frac{-t}%
{4m^{2}-t}}~\mathcal{R}^{2}(t),
\end{equation}%
\begin{equation}
\left(  T_{2}+T_{4}\right)  \bar{T}_{5}^{\ast}\sim \mathcal{O}(\alpha_{s}) \approx 0.
\label{T2pT4T5}
\end{equation}
Using  Eqs.(\ref{T1-T3}), (\ref{T13=SG1}) and (\ref{T5=C5G1})  we also obtain
\begin{equation}
(\bar{T}_{3}-\bar{T}_{1})T_{6}^{\ast}\simeq m\Delta ~C_{6}{\frac{-t}{4m^{2}-t}}~G_{0}(t)\mathcal{R}(t),
\end{equation}%
\begin{equation}
(\bar{T}_{3}+\bar{T}_{1}){T}_{5}^{\ast}\simeq m^{2}\Sigma ~C_{5}\sqrt{\frac{-t}{4m^{2}-t}} ~G_{1}(t)\tilde{G}_{1}(t).
\label{T3pT1T5}
\end{equation}
From the  given expressions one can easily observe that  the contribution proportional to  $r_{0}$  appears in the numerator of all asymmetries and therefore one can expect that these observables can be more sensitive to this subleading amplitude.   By evaluating  these asymmetries at $-t=2.5$~GeV$^{2}$,  we obtain 
\bea
\KLL[s=9~\text{GeV}^{2},\theta =71.5^{o}]&=&\frac{0.46+0.27 r_{0}+0.01r_{1}\tilde{r}_{1} }{1.47+3.1 r_{0}^{2}+0.03 r_{1}^{2}+0.007~\tilde{r}_{1}^{2}}\approx
\frac{0.46+0.27 r_{0}}{1.47+3.1 r_{0}^{2}}, \label{KLL25}
\\[2mm]
\KLS[s=9~\text{GeV}^{2},\theta =71.5^{o}]&=&\frac{0.36-0.34 r_{0}+0.009 r_{1}\tilde{r}_{1} }{1.47+3.1 r_{0}^{2}+0.03 r_{1}^{2}+0.007~\tilde{r}_{1}^{2}}\approx
\frac{0.36-0.34 r_{0}}{1.47+3.1 r_{0}^{2}}, \label{KLS25}
\\[2mm]
\ALL[s=8~\text{GeV}^{2},\theta =78^{o}]&=&\frac{0.56+0.28 r_{0}-0.02 r_{1}\tilde{r}_{1} }{1.49+2.7 r_{0}^{2}+0.06 r_{1}^{2}+0.01~\tilde{r}_{1}^{2}}\approx
\frac{0.56+0.28 r_{0}}{1.49+2.7 r_{0}^{2}}. \label{ALL25}
\eea
We again observe that the contributions proportional to  $r_{1}$ and $\tilde r_{1}$ are practically negligible.  In this case,  all three asymmetries depend on the same unknown quantity $r_{0}$ at fixed momentum transfer.   Assuming that  $r_{0}$ is restricted as in (\ref{r0est}) we find
\bea
\KLL[s=9,\theta =71.5^{o}]=0.31^{+0.01}_{-0.04},
\label{KLL9}
\eea
where the central number is computed at $r_{0}=0$. The  uncetainty in (\ref{KLL9}) is  smaller than the estimated  statistical accuracy $\pm 0.06$ in this experiment \cite{E07002}.   It is natural to expect that  $\KLS$ is more sensitive to  the value $r_{0}$ because in this observable  helicity-flip  contributions are not power suppressed.   Using (\ref{KLS25}) we find
 \bea
\KLS[s=9,\theta =71.5^{o}]=0.24^{+0.03}_{-0.04}
\eea
yielding an uncertainty of around $16\%$ which is  smaller  than the expected   statistical accuracy $\pm 0.05$  \cite{E07002}, for preliminary result, see Ref.\cite{Fanelli:2014yda}.  

If we  assume  that in the leading-order  approximation the  combination $\left(  T_{2}+T_{4}\right)  \bar{T}_{5}^{\ast}$  is small, see Eq.(\ref{T2pT4T5}),  then the analytical expressions for the two asymmetries $\KLL$ and $\ALL$  only  differ by the combination in Eq.(\ref{T3pT1T5})  $(\bar{T}_{3}+\bar{T}_{1}){T}_{5}^{\ast}\sim r_{1}\tilde r_{1}$.  But as one can observe from Eqs.(\ref{KLL25}) and (\ref{ALL25}) that the corresponding contribution is  numerically small and therefore  one obtains that   $\KLL\simeq \ALL$.  The uncertainty provided by the ratio $r_{0}$  in $\ALL$  in Eq.(\ref{ALL25}) yields:
\bea
\ALL[s=8,\theta =78^{o}] = 0.37^{+0.02}_{-0.04},
\label{ALL8}
\eea    
around  $11\%$  which is again smaller than  the statistical accuracy discussed in Ref.\cite{PR1214006}.  

\begin{figure}[h]
\centering
\includegraphics[height=2.0955in]{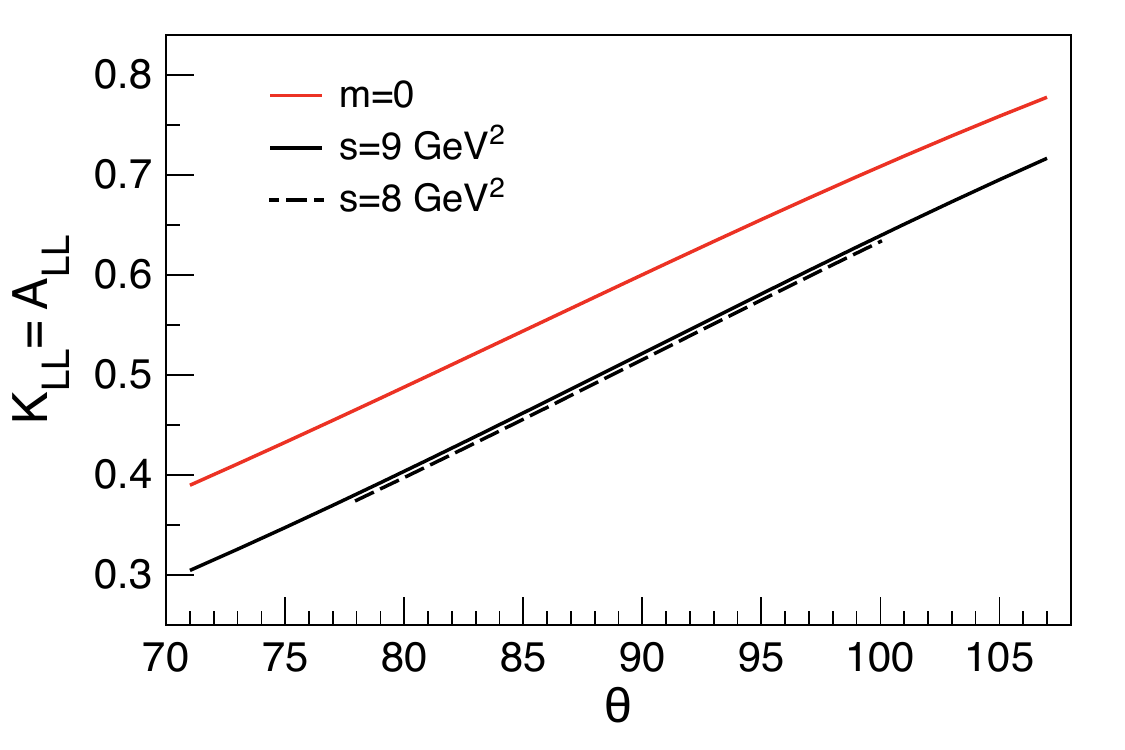}
\caption{The asymmetry KLL as a function of scattering angle. The red solid curve shows the
result without  kinematical power corrections (Klein-Nishina result). The black dashed and solid lines correspond to $s=8,9$~GeV$^{2}$, respectively. }
\label{kllkin8-9}
\end{figure}
In order to see the effect of the kinematical power corrections we plot in Fig.\ref{kllkin8-9}  asymmetry $\KLL$  as a function of scattering angle for two different values of  energy $s=8,9$~GeV$^{2}$  with and without power suppressed contributions.  All helicity-flip contributions are taken to be zero $r_{0}=r_{1}=\tilde r_{1}=0$.   The red line in Fig.\ref{kllkin8-9} denotes the asymmetry without the power corrections which  reduces to the Klein-Nishina  result  on the pointlike massless target $\KLL^{KN}=(4-(1+\cos\theta)^{2})/(4+(1+\cos\theta)^{2})$.  We only consider  the angles for which $-t,-u\geq 2.5$~GeV$^{2}$.  In this region   the power corrections do not change the angular dependence  but reduce the value of the massless asymmetry by $25\%$.  One can also observe that the values of $\KLL$ at   both values of $s$ are almost the same. This prediction can be checked   by measuring the asymmetry  $\ALL$ in the new experiment \cite{PR1214006}  at the same angles as $\KLL$ measured  in \cite{E07002}.   

\section{Discussion}
\label{dis}

In this work we presented a phenomenological analysis of  the cross section and asymmetries of wide angle Compton scattering in which
we accounted for   different power suppressed contributions. 
For the first time we include in the analysis the subleading helicity-flip amplitudes using  the SCET framework.
We assume that the dominant contribution to these amplitudes is provided 
by the soft-overlap configurations  described by the matrix elements of
 SCET-I  operators. We only consider the operators which  appear  in the leading-order approximation. 
The corresponding hard-coefficient functions were also computed.  Within this formalism we estimated  the effect  due to 
 the power suppressed corrections  in different WACS observables . 
     
An analysis of  existing cross section data allows us to conclude that the developed description can work reasonably well in the region where
 $-t,-u>2.5$~GeV$^{2}$. The contribution  from the helicity-flip amplitudes in the cross section  is smaller than  $10\%$.  We also found  that
 the corresponding  effect due to power corrections   in the different asymmetries are also relatively small and to a good accuracy $\ALL = \KLL$ 
 in the relevant kinematical region.        
 
 \section{Acknowledgements}
 
 This work is supported by Helmholtz Institute Mainz.  N.K. is also grateful to German-US exchange program on Hadron Physics for financial support and to  staff of  Old Dominion University for warm hospitality during his visit.



\begin{thebibliography}{9}  

\bibitem{Shupe:1979vg}
  M.~A.~Shupe, R.~H.~Milburn, D.~J.~Quinn, J.~P.~Rutherfoord, A.~R.~Stottlemyer, S.~S.~Hertzbach, R.~RKofler and F.~D.~Lomanno {\it et al.},
  Phys.\ Rev.\ D {\bf 19} (1979) 1921.
  
\bibitem {Danagoulian:2007gs}A.~Danagoulian \textit{et al.} [Hall A
Collaboration],
Phys.\ Rev.\ Lett.\ \textbf{98} (2007) 152001 [nucl-ex/0701068 [NUCL-EX]].


\bibitem {Hamilton:2004fq}D.~J.~Hamilton \textit{et al.} [Jefferson Lab Hall A
Collaboration],
Phys.\ Rev.\ Lett.\ \textbf{94} (2005) 242001 [nucl-ex/0410001].

\bibitem{PR1214003}, D.~Hamilton \textit{et al.} [Jefferson Lab Hall C
Collaboration], E1214003,
``Wide-angle Compton Scattering at 8 and 10 GeV Photon Energies''. 
 

\bibitem{Kronfeld:1991kp}
  A.~S.~Kronfeld and B.~Nizic,
  Phys.\ Rev.\ D {\bf 44} (1991) 3445
   [Erratum-ibid.\ D {\bf 46} (1992) 2272].
  
\bibitem{Vanderhaeghen:1997my}
  M.~Vanderhaeghen, P.~A.~M.~Guichon and J.~Van de Wiele,
  Nucl.\ Phys.\ A {\bf 622} (1997) 144C.
  
\bibitem{Brooks:2000nb}
  T.~C.~Brooks and L.~J.~Dixon,
  Phys.\ Rev.\ D {\bf 62} (2000) 114021
  [hep-ph/0004143].
  
\bibitem{Thomson:2006ny}
  R.~Thomson, A.~Pang and C.~-R.~Ji,
  Phys.\ Rev.\ D {\bf 73} (2006) 054023
  [hep-ph/0602164].




\bibitem{Radyushkin:1998rt}
  A.~V.~Radyushkin,
  Phys.\ Rev.\ D {\bf 58} (1998) 114008
  [hep-ph/9803316].
  
\bibitem{handbag1}
  M.~Diehl, T.~Feldmann, R.~Jakob and P.~Kroll,
  Eur.\ Phys.\ J.\ C {\bf 8} (1999) 409
  [hep-ph/9811253].
  \bibitem{handbag2}
  M.~Diehl, T.~Feldmann, R.~Jakob and P.~Kroll,
  Phys.\ Lett.\ B {\bf 460} (1999) 204
  [hep-ph/9903268].
  \bibitem{handbag3}
  M.~Diehl, T.~Feldmann, H.~W.~Huang and P.~Kroll,
  Phys.\ Rev.\ D {\bf 67} (2003) 037502
  [hep-ph/0212138].

\bibitem{Kroll:1990kh}
  P.~Kroll, M.~Schurmann and W.~Schweiger,
  Int.\ J.\ Mod.\ Phys.\ A {\bf 6} (1991) 4107.
  
\bibitem{Miller:2004rc}
  G.~A.~Miller,
  Phys.\ Rev.\ C {\bf 69} (2004) 052201
  [nucl-th/0402092].

\bibitem{Kivel:2012vs}
  N.~Kivel and M.~Vanderhaeghen,
  JHEP {\bf 1304} (2013) 029
  [arXiv:1212.0683 [hep-ph]].

\bibitem{Kivel:2013sya}
  N.~Kivel and M.~Vanderhaeghen,
  Nucl.\ Phys.\ B {\bf 883} (2014) 224
  [arXiv:1312.5456 [hep-ph]].
  
  
\bibitem {Babusci:1998ww} D.~Babusci, G.~Giordano, A.~I.~L'vov, G.~Matone and
A.~M.~Nathan,
Phys.\ Rev.\ C \textbf{58} (1998) 1013  [hep-ph/9803347].

 

\bibitem{Bauer:2000ew} C.~W.~Bauer, S.~Fleming and M.~E.~Luke,
Phys.\ Rev.\ D \textbf{63}, 014006 (2000).


\bibitem{Bauer2000}C.~W.~Bauer, S.~Fleming, D.~Pirjol and I.~W.~Stewart,
Phys.\ Rev.\ D \textbf{63}, 114020 (2001).


\bibitem {Bauer:2001ct}C.~W.~Bauer and I.~W.~Stewart,
Phys.\ Lett.\ B \textbf{516}, 134 (2001).


\bibitem {Bauer2001}C.~W.~Bauer, D.~Pirjol and I.~W.~Stewart,
Phys.\ Rev.\ D \textbf{65}, 054022 (2002).


\bibitem{Beneke:2002ph}
M.~Beneke, A.~P.~Chapovsky, M.~Diehl and T.~Feldmann,
Nucl.\ Phys.\ B \textbf{643}, 431 (2002).


\bibitem{Beneke:2002ni}
M.~Beneke and T.~Feldmann,
Phys.\ Lett.\ B \textbf{553}, 267 (2003).


 
\bibitem{Belitsky:2002kj}
  A.~V.~Belitsky, X.~d.~Ji and F.~Yuan,
  Phys.\ Rev.\ Lett.\  {\bf 91} (2003) 092003
  [hep-ph/0212351].
  
  
\bibitem{Kivel:2010ns}
  N.~Kivel and M.~Vanderhaeghen,
  Phys.\ Rev.\ D {\bf 83} (2011) 093005
  [arXiv:1010.5314 [hep-ph]].
  
\bibitem{Jones:1999rz}
  M.~K.~Jones {\it et al.}  [Jefferson Lab Hall A Collaboration],
  Phys.\ Rev.\ Lett.\  {\bf 84} (2000) 1398
  [nucl-ex/9910005].

\bibitem{Punjabi:2005wq}
  V.~Punjabi, C.~F.~Perdrisat, K.~A.~Aniol, F.~T.~Baker, J.~Berthot, P.~Y.~Bertin, W.~Bertozzi and A.~Besson {\it et al.},
  Phys.\ Rev.\ C {\bf 71} (2005) 055202
   [Erratum-ibid.\ C {\bf 71} (2005) 069902]
  [nucl-ex/0501018].
  
\bibitem{Meziane:2010xc}
  M.~Meziane {\it et al.}  [GEp2gamma Collaboration],
  Phys.\ Rev.\ Lett.\  {\bf 106} (2011) 132501
  [arXiv:1012.0339 [nucl-ex]].


\bibitem{E07002}, P.~Bosted  \textit{et al.}  [Jefferson Lab Hall C
Collaboration], E-07-002,
``Polarization transfer in Wide Angle Compton scattering''. 

\bibitem{Fanelli:2014yda} 
  C.~Fanelli, E.~Cisbani, D.~Hamilton, G.~Salme and B.~Wojtsekhowski,
  EPJ Web Conf.\  {\bf 66}, 06006 (2014).


\bibitem{PR1214006}, D.~Nikolenko   \textit{et al.}  [Jefferson Lab Hall C
Collaboration], PR-12-14-006,
``Initial State Helicity Correlation in Wide Angle Compton scattering''. 

    
 
\end{thebibliography}
\end{document}